\documentclass[conference]{IEEEtran}
\IEEEoverridecommandlockouts
\usepackage{amsmath,amssymb,amsfonts}
\usepackage{algorithmic}
\usepackage{graphicx}
\usepackage{textcomp}
\usepackage{listings}
\usepackage{xcolor}
\usepackage{hyperref}
\usepackage{biblatex}
\usepackage[disable]{todonotes}

\definecolor{javared}{rgb}{0.6,0,0} 
\definecolor{javagreen}{rgb}{0.25,0.5,0.35} 
\definecolor{javapurple}{rgb}{0.5,0,0.35} 
\definecolor{javadocblue}{rgb}{0.25,0.35,0.75} 
\definecolor{gray}{rgb}{0.5,0.5,0.5}

\lstdefinestyle{customjava}{
	language=Java,
	breaklines=true,
	showstringspaces=false,
	basicstyle=\linespread{1}\fontfamily{fvm}\selectfont,
	keywordstyle=\color{javapurple}\bfseries,
	commentstyle=\color{javagreen}\itshape,
	stringstyle=\color{javared},
	columns=flexible,
	tabsize=2,
	showspaces=false,
	showstringspaces=false,
	escapeinside={(*@}{@*)},
	escapechar=~,
	numbers=left,
	stepnumber=1,
	numberstyle=\small,
	numbersep=10pt,
	xleftmargin=2em,
	frame=single,
	framexleftmargin=1.5em,
	framerule=0pt
}

\newcommand{\javaCode}[1]{\lstinline[style=customjava]{#1}}
\lstnewenvironment{java}{\lstset{style=customjava}}{}
\iffalse
\usepackage{censor}
\newcommand{\censorCite}[1]{\censor{[10]}}
\else
\newcommand{\censorCite}[1]{\cite{#1}}
\newcommand{\censor}[1]{#1}
\fi

\if false
----------------------- REVIEW 1 ---------------------
SUBMISSION: 1769
TITLE: A Feedback Toolkit and Procedural Guidance for Teaching Thorough Testing
AUTHORS: Steffen Dick, Christoph Bockisch, Harrie Passier, Lex Bijlsma and Ruurd Kuiper

----------- Overall evaluation -----------
SCORE: 0 (borderline paper)
----- TEXT:
The authors present a novel approach to check and subsequently provide automated
feedback for code submissions of students.

They present a tool-extension for an e-learning platform that supports the
definition of a teachers solution of an assignment

and allows to define code-coverage related, predefined and condition-based feedback
text to a student’s solution of a unit-test covering the assignment’s goals.

PROBLEM: Due to the included link in section V (https://t.ly/b0V6) a possible
conflict with the guidelines for a double-blind review might be violated.

The PROs of the paper are:

               - easy to follow examples which motivate the application of the method

               - a more extensive example provided online

The CONs of the paper are:

               - that the point being made is sometimes difficult to follow due to
the structure of and how the text is presented

               - previous work (maybe of the authors) is taken as a given and often
referenced as a broader concept.

                              This makes it difficult for the reader to understand
the contribution of the article.

               - The method is intended to be used mainly for a specific software
platform (Quaterfall) and language (Java).

                              The authors lack to outline how the concept can be
generalized to other tools/platforms

The core idea to use annotated reference code to provide guided feedback for
unit-tests is novel.

Yet the authors themselves state in the abstract that "...developed a toolkit that
students can use to independently get individual feedback on their solution and the
adequateness of their tests."

is problematic as the method does not address generic code input, but rather
provides hints the teacher thought of in advance.

This does not necc. support the stated goal to improve the "Correctness of software
[..] one of the more important criteria of qualitative software".

The technical quality is fair. The examples given (including the online resource)
are sound.

The bibliography is very compact and some text passages with a subsequent citation
are difficult to link to the reference they point to.

Comments:

               - the novelty of the article should be made more clear compared to
existing tools (as given in the referenced work) and classical
coverage reports.

                              What are disadvantages for the teacher/students?

                              What is the added benefit for the teacher? (currently
it looks like overhead to set up a solution + config
in advance)

               - Limitations of the work should be addressed

                              What happens in the case of complex assignments where
the implementation can be done in different ways
(e.g., loop vs. recursion)?

                              How can concepts of OOP be addressed(e.g.,
private/public methods, inheritance...)?

                              Which languages can be addressed? (the authors state
that this approach could be ported to other
tools/languages)          

               - the text contains some grammar issues that can be easily fixed with
a proof-read

               - the text contains some phrases that should be avoided in a
scientific paper (e.g., "..as we will see in a bit.")

               - an attempt should be made to improve the readability of the text
with regard to the aim of the paper, separating preliminaries and
actual contributions more clearly

               - it should be made more clear in the introduction that the following
work is intended for Java and Quaterfall (currently part of section
3)

                              Additionally, the focus of the presented work should
be highlighted more and contrasted to previous work /
how it is addresses the often referenced procedural
guidance approach

               - the article references a broader concept in section II (referring
to an extensive report) which serves as a baseline for the work.

                              An effort should be made to make this more
comprehensible with respect to the goal of the
presented work.

               - in the given code examples, the definition of the class "Person"
has missing content: "... information about a, ..."

                              Is there missing text due to the line break - Or is
this done intentionally?

               - a small figure/ textual aggregation to provide a compact overview
of the MASS-Toolkit would be helpful

----------------------- REVIEW 2 ---------------------
SUBMISSION: 1769
TITLE: A Feedback Toolkit and Procedural Guidance for Teaching Thorough Testing
AUTHORS: Steffen Dick, Christoph Bockisch, Harrie Passier, Lex Bijlsma and Ruurd Kuiper

----------- Overall evaluation -----------
SCORE: -1 (weak reject)
----- TEXT:
Summary
This paper describes a process for teaching students how to program along with tool
support for that method. The automated tool provides tailorable feedback at various
stages along the way to help students learn. The benefit of this approach is that it
helps the students determine what is missing rather than just giving them the
correct answer.

Overall Comments
The idea of this tool is interesting. I think it is reasonable to tailor the
feedback to help the students learn rather than to point them directly to the
missing information. From what I can tell in the paper, the tool seems to be
appropriate to support the pedagogy.

However, there is no data in the paper to provide evidence of its usefulness. At one
point, the authors discuss a small number of improvements that need to be made based
on anecdotal feedback from students. For anyone else to find this tool useful and be
willing to invest the effort to use it, there needs to be an evaluation. At this
point, the paper is premature until there is solid data that shows the tool is
actually useful based on some predefined measures and research goals.

The authors also miss a number of related works that are similar to what is being
proposed here. Granted some of these are older, so the authors should consider their
relevance and potentially look at other follow-on work. For example:
- J. Collofello and K. Vehathiri. 2005. An Environment for Training Computer Science
Students on Software Testing. Proceedings Frontiers in Education 35th Annual
Conference. https://doi.org/10.1109/fie.2005.1611937
- D. M. de Souza, B. H. Oliveira, J. C. Maldonado, S. R. S. Souza, and E. F.
Barbosa. 2014. Towards the use of an automatic assessment system in the teaching of
software testing. In 2014 IEEE Frontiers in Education Conference (FIE) Proceedings.
1–8.
- Lucas Cordova, Jeffrey Carver, Noah Gershmel, and Gursimran Walia. 2021. A
Comparison of Inquiry-Based Conceptual Feedback vs. Traditional Detailed Feedback
Mechanisms in Software Testing Education: An Empirical Investigation. In Proceedings
of the 52nd ACM Technical Symposium on Computer Science Education (SIGCSE '21).
Association for Computing Machinery, New York, NY, USA, 87–93.
https://doi.org/10.1145/3408877.3432417
- Stephen H. Edwards. 2004. Using software testing to move students from trial-and
error to reflection-in-action. Proceedings of the 35th SIGCSE technical symposium on
Computer science education - SIGCSE 04 (May 2004), 26.
https://doi.org/10.1145/971300.971312
- Jaime Spacco, David Hovemeyer, William Pugh, Fawzi Emad, Jeffrey K. Hollingsworth,
and Nelson Padua-Perez. 2006. Experiences with marmoset. Proceedings of the 11th
annual SIGCSE conference on Innovation and technology in
computer science education - ITICSE 06 (2006). https://doi.org/10.1145/1140124.1140131

----------------------- REVIEW 3 ---------------------
SUBMISSION: 1769
TITLE: A Feedback Toolkit and Procedural Guidance for Teaching Thorough Testing
AUTHORS: Steffen Dick, Christoph Bockisch, Harrie Passier, Lex Bijlsma and Ruurd Kuiper

----------- Overall evaluation -----------
SCORE: -1 (weak reject)
----- TEXT:
The authors pointed out a problem students do not receive sufficient feedback on
code quality in programming education. 
To solve this problem, this study developed a procedural guidance and developed a
toolkit.

It is important and interesting.

This paper does not show evidences of effectiveness of the approach.
\fi

\begin{document}


\title{A Feedback Toolkit and Procedural Guidance for Teaching Thorough Testing
 \thanks{This work is partly funded by the \censor{Erasmus+} project \censor{\textit{Quality-focused Programming Education (QPED)}}, \censor{2020-1-NL01-KA203-064626.}}
}

 \author{\IEEEauthorblockN{\censor{Steffen Dick \quad Christoph Bockisch}}
 \IEEEauthorblockA{\censor{\textit{Fachbereich 12 Mathematik und Informatik}} \\
 \censor{\textit{Philipps-Universität Marburg, Germany}}\\
 \censor{35032 Marburg, Germany} \\
 \censor{\{dickst,bockisch\}@mathematik.uni-marburg.de}}
 \and
 \IEEEauthorblockN{\censor{Harrie Passier \quad Lex Bijlsma \quad Ruurd Kuiper}}
 \IEEEauthorblockA{\censor{\textit{Faculty of Science, Department of Computer Science}} \\
 \censor{\textit{Open Universiteit, The Netherlands}}\\
 \censor{6401 DL Heerlen, The Netherlands} \\
 \censor{\{harrie.passier,lex.bijlsma,ruurd.kuiper\}@ou.nl}}
 }

\maketitle

\begin{abstract}
Correctness is one of the more important criteria of qualitative software.
However, it is often taught in isolation and most students consider it only as an afterthought.
They also do not receive sufficient feedback on code quality and tests unless specified in the assignment.
To improve this, we developed a procedural guidance that guides students to an implementation with appropriate tests.
Furthermore, we have developed a toolkit that students can use to independently get individual feedback on their solution and the adequateness of their tests.
A key instrument is a test coverage analysis which allows for teachers to customize the feedback with constructive instructions specific to the current assignment to improve a student's test suite.
In this paper, we outline the procedural guidance, explain the working of the feedback toolkit and present a method for using the toolkit in conjunction with the different steps of the procedural guidance.
\end{abstract}



\section{Introduction}
Programming education is an integral part of computer science and related subjects.
Often, graduates of these studies will be developing software in their job.
Therefore, it is essential to teach them to write high-quality and correct software.
Even though we mention the importance of writing clean code and testing in our courses, our teaching usually focuses on the result rather then the way to get there.
For example, teaching, especially in early programming courses, is often organized along features of a programming language or the capabilities of libraries such as JUnit (for testing) rather than the means for employing these features systematically.
Similarly, when grading solutions to an assignment, we usually assess its functional correctness or the technically correct usage of language concepts and rarely how students reached their solution.
While we expect that students test the solution, these tests are often not part of the assignment and are therefore not handed in or assessed by the teacher.
As a result, students are not well trained in systematically developing the functional and corresponding testing code.

To improve this, we firstly developed a procedural guidance reaching from the analysis of a problem to a well-tested, high-quality implementation.
The guidance structures the development process into its main activities and provides explicit procedure-like rules to order the steps in the activities.
This general procedure can be mirrored in specific assignments used in the course, e.g., in the problem description and the way that sub-questions lead students through applying the procedure.

It is important to give students quick feedback on their activities.
To make this feasible, especially in case where the student per staff ratio is high and in online teaching, tool support is essential.
Several tools exist that provide such feedback.
However, such tools are often difficult to use for students in the beginning of their studies, specifically with respect to feedback on testing.
Nevertheless, Cavalcanti et al. found improvement in student performance when automatic feedback was introduced \cite{CAVALCANTI2021}, which encourages further investigation.
So, secondly, we tie tool support for automatic feedback to our procedural guidance.
We provide feedback on wheter students pay sufficient attention to the procedure and on the quality of the artefacts produced.
In particular, we address the often neglected problem of the quality of tests.

To improve on the general automatic feedback issue, we have developed a suite of feedback tools which we call MASS.
Existing tools are integrated into this suite and feedback can be customized to the specific assignments the students may face.
The different components, called checkers, analyse different aspects of code submitted by students.
For some of these aspects, MASS has general rules, where teachers can configure the systems' strictness or adapt rules---as required by the standards that the procedural guidance aims for.
For other aspects, teachers can configure messages more individually on a per-assignment basis---connecting to the procedural guidance.
MASS provides students and teachers with an accessible, 24/7 available tool set that gives understandable feedback and relates to the procedural guidance approach at various levels.

In this paper, we focus on the test-coverage checker we have developed.
We show how the procedural guidance can be used to support the testing.
Using artefacts from the procedural guidance, teachers provide instructional texts for the automatic feedback that guides students to identify, e.g., the missing test cases for a specific assignment and improve their tests.

We briefly present the procedural guidance and its teaching approach in general in Section II. In Section III, we introduce the MASS feedback toolkit in general.
The test coverage checker is described in Section IV. We then demonstrate the combined usage of the procedural guidance together with MASS by means of a detailed, fully worked out, online accessible example in Section V.
Finally, in Section VI we present our experiences with using the procedural guidance and the MASS toolkit.
Lastly, in Section VII, we outline future improvements based on these experiences.

\section{Procedural Guidance}
Programming education should educate students to be able to develop good quality software.
This is a complex activity \cite{merrienboer2007} in which many decisions have to be made about many different issues.
Besides the conceptual knowledge, procedural knowledge plays an important role during this process.
The former is well-served by current teaching practices and materials, the latter less so.
Procedural guidance is a powerful approach to teach procedural knowledge by providing explicit step-by-step guidance to perform steps in complex activities.

We provide procedural guidance for the development of object-oriented software, which is fully elaborated in a technical report \censorCite{passier2022}.
In that approach, classes annotated with a specification are the unit of development.
The procedural guidance concerns two levels of detail.

The first level of guidance structures the activity of the student.
In order to develop a class, three views are distinguished, each resulting in an artefact: the \textit{External View} (also known as API), the \textit{Internal View} (concerning attributes and private methods), and the \textit{Annotated Code View} (the programming code itself).
Each of these artefacts is developed in three main activities, namely \emph{Analysis} (collecting all the information needed to produce the design and specification), \emph{Design} (in which the structure is determined), and S\emph{pecification} (in which the behaviour is described).

Additional to the development of a class itself, tests are developed.
For each of the views corresponding tests are developed: External (Blackbox) tests, Internal (Greybox) tests, and Code (Whitebox) tests.
This amounts to translation of the specifications from the views into test cases, which is relatively straightforward.
Each method M has several specifications C. For each combination `method M - specification C' a test method with the name \texttt{testMC} is defined.
The test method's assertions follow from the specification.
This way, the connection between specification parts and tests is made.
Therefore, the Analysis, Design and Specification activities for tests are combined into the activity \emph{Construction}.
\begin{figure}
    \centering
    \includegraphics[width=\linewidth]{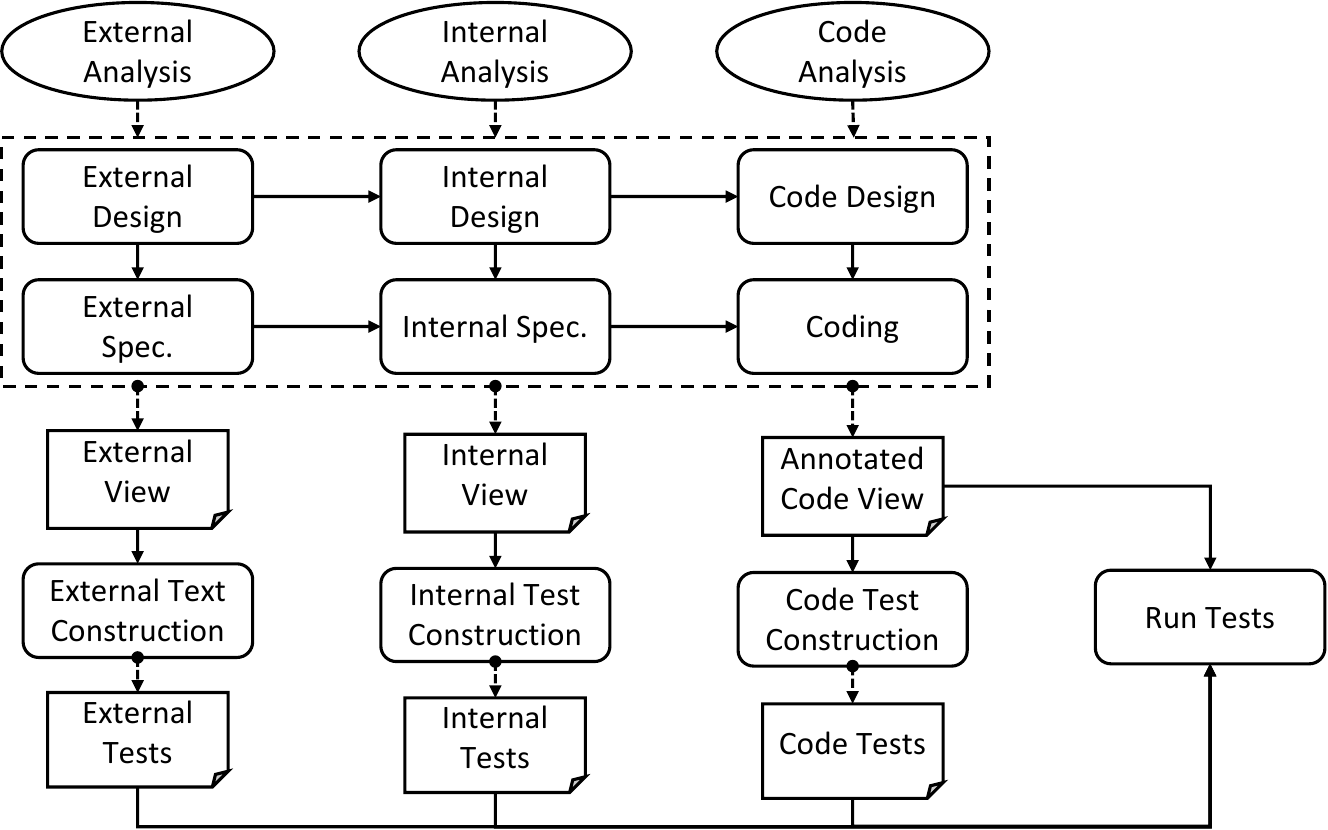}
    \caption{Overview of the procedural guidance}
    \label{fig:pgoverview}
\end{figure}

Figure \ref{fig:pgoverview}, from the technical report, provides an overview of this structure.
The relationships are more precisely argued and defined in the technical report, for the present paper, however, intuition suffices.
Since the Procedural Guidance is flexible, e.g. with respect the order of of steps and their repetition, it can be used to support different development methods and didactic approaches.

As an example (further elaborated in Section  V), think of developing a program that implements, possibly as part of a larger program, a class \texttt{Bag}. This is a collection where, different form sets, duplicate elements are supported.
\begin{itemize}
    \item The External View artefact would be declarations (i.e., partial code) and specifications of public methods like \texttt{add}, \texttt{remove} and \texttt{display}.
    \item The Internal View artefact would be extensions of Eternal View with instance variables like a---possibly sorted---array (or a linked list, etc.) to store the elements and a private method like \texttt{insert}.
    \item The Code View would be the extension of the Internal View to a complete, annotated, implementation, like implementing the sorting.
\end{itemize}
The second level of guidance is rule-like, step-wise guidance as to how to perform steps in the various activities that produce the artefacts.
This guidance is therefore given as a set of rules grouped with the activities.

For the example, think of general rules how to choose a sensible external/internal interface, how to decide on instance variables, how to maintain and argue consistency between the artefacts, how to decide on tests, ensure test coverage, execute and interpret tests, and so on.

The procedural guidance is general. To help the student understand and achieve proficiency in the procedural guidance, assignments can be used.
Such specific steering can obviously not be part of the general approach.
However, the general structuring and the step-wise guidance greatly help the provider of an assignment (whether it be a teacher or a contributor to an assignment repository) to devise such steering.
This helps the student to understand and apply the general procedural guidance rules.
Such guidance may well take into account the progress of a student:
It may range from providing (parts of) artefacts to novices up to not providing any advice to advanced students.
Similar options to tie procedural guidance to assignments exist in the feedback.

For the example, think of \emph{supplying} a complete External View and some External tests to novices, or \emph{hinting} an intermediate student at the use a (sorted) array for implementation and at testing for it being empty or overflowing.

Some remarks:
\begin{itemize}
    \item The approach not only supports development for correctness of core functionality, but also for robustness:
    after ensuring correct behavior on the `happy path' (i.e., when the software is used according to the specification), all artefacts are extended to ensure sensible behaviros on the `non-happy path' following the same procedures.
    \item The approach is flexible in the level of formality.
    The specification uses part of JML \cite{burdy2003} tags and is inspired by the JML notation.
    Since the specification is mainly intended to be read by humans, natural language and JML specifications are allowed to be mixed.
    \item Ultimately, the student should become accustomed to the practices targeted by the procedural guidance such that the guidance will no longer be needed as scaffolding.
\end{itemize}

\section{The MASS toolkit}
MASS, \censor{the Marburg University auto-assessment system}, is comprised of a variety of different tools meant to be used to give students understandable, automatically generated feedback on their programming exercises.
MASS was originally designed to be used in conjunction with the Quarterfall platform\footnote{The Quarterfall homepage: \url{https://www.quarterfall.com}}, which is an e-learning platform allowing teachers to offer learning material including assignments and students to process assignments and submit solutions.
The platform offers a built-in mechanism for generating feedback on student answers by using unit tests provided by teachers which makes it comparable to other such platforms as it only tests on functional correctness.
However, Quarterfall offers \emph{Cloud Checks} which allows the platform to integrate tools for checking student answers in more sophisticated ways.
The communication between the platform and the plug-in is done via a JSON-object that Quarterfall provides to the plug-in.
Because MASS only uses a JSON-object to communicate with a given platform, it can be easily integrated into other platforms that can provide a JSON-object with all the necessary information.
The full source code of MASS is available at our open-source Git repository. \censor{\url{https://github.com/qped-eu/MASS-checker}}.

In MASS, we combined several feedback tools that can be configured individually and at a fine granularity.
The selected tools are all used for the Java programming language, but similar tools are available for all other major programming languages.
Each \emph{checker} can assess code and generate feedback for the given scenario by analyzing a different aspect of the code, either by means of static or dynamic analysis.
Without going into too much detail, we will briefly discuss the different checkers and their uses.

\begin{itemize}
    \item The \emph{Syntax-Checker} is used to assess the syntax and other language constraints (such as typing) of handed-in code.
    It generally generates feedback that is more suitable to beginners than the feedback usually given by the Java compiler.
    The feedback itself can be customized using two different levels: beginner and advanced.
    \item The \emph{Style-Checker} can assess code regarding violations of styling guidelines (e.g. the official Java guidelines).
    Its feedback can be customized by using regular expressions and proficiency levels.
    \item The \emph{Solution-Approach-Checker} can be used to assess the way a student solved a given problem.
    This can, for example, be customized to detect use of recursion or limit the number and kind of loops used within the solution.
    \item The \emph{Class-Info-Checker} assesses the design of classes within a given project.
    It gives feedback on different access modifiers, the inheritance of the class or, for instance, expected super classes.
    \item The \emph{Metrics-Checker} uses different object-oriented metrics to assess the quality of a given project.
    Feedback can be specifically tailored to the metrics chosen and customized by the teacher.
     \item The \emph{Test-Coverage-Checker} is the main piece of this paper and will be discussed in detail in the next section.
\end{itemize}

\section{Test-Coverage-Checker}
The purpose of the test coverage checker is to give students automatically generated feedback on the completeness of their tests.
This is achieved by running the tests contained within a solution submitted by the student, monitoring the lines of code that were covered or not covered during the execution of the tests, and producing feedback based on the gathered coverage information.
For our implementation for Java, the test coverage checker internally runs tests with JUnit (Jupiter) and uses the JaCoCo test coverage tool.
The checker can run the tests within the submissions of students either against the code within the same submission or against a previously configured private implementation of the code by the teacher.
A private implementation can either fully replace the student implementation or replace parts thereof.
This allows for providing very specific and individual feedback messages.
It is also possible for the private implementation to include tests.
With this, the functional correctness of student solutions can also be checked with the same tool, which we will exploit in the integrated example in Section V.

To discuss the capabilities and the configuration of the test coverage checker, we will use a simple example assignment:
\emph{Write tests for a method \javaCode{boolean isEven(int n)} that returns  \javaCode{true} if a given number n is an even number, and  \javaCode{false} otherwise.}
Figure \ref{fig:evenImplementation} shows an implementation of the method which is not provided to students and will be used as a private implementation.

\begin{figure}
    \centering
    \begin{java}
class Even {
  public boolean isTrue(int num) {
    if (num % 2 == 0) {
      return true;
    } else {
      return false;
    }
  }
}
\end{java}
    \caption{The implementation of \texttt{Even}}
    \label{fig:evenImplementation}
\end{figure}

The teacher knows the implementation against which students will run tests and even knows the exact meaning of each line of code.
Thus, the teacher can configure specific messages that are triggered when certain lines are not reached within the code.
For example, if line 4 is not reached, one such message can be: \emph{``You should test for odd numbers as well.''}
Since the message is custom-provided, we can also choose to make it more directive (e.g., \emph{``Add a test case for \texttt{num = 2}''} or less directive (e.g., \emph{``You did not consider all cases for the parity of numbers''}), depending on the students' experience and the didactic approach of the teacher.
Of course, there would be an analogous feedback message for the case that line 6 is missed.

Also, in particular with very directive feedback messages, one may come to the conclusion that a line-specific message might end up giving away too much:
Consider a submission where the student does not write any tests, waits for the feedback and then simply implements the tests according to what the feedback entails.
To prevent this behaviour, it is possible to configure a hierarchy of messages where a higher ranked message can suppress a lower ranked message.
In our example, this means that a higher ranked message should be implemented for line 3 that suppresses both messages for line 4 and 6.
This means that students would only get a message that directs them to test the method instead of giving away the solution of testing even and odd numbers.

\subsection{Configuration}
The configuration can entirely be done within a JSON-object, which can be hand written or generated with the configurator application on our web page.
\footnote{\censor{Mass Configurator: \url{https://qped-eu.github.io/mass/\#/configurator}}}
In both cases, it is necessary to know and understand the configuration properties and their possible values.
For visual clarity, we will focus on the use of the website instead of the JSON-object.
A full documentation of the MASS toolkit, including the schema for the JSON-object, can be found online.
\footnote{\censor{MASS configuration documentation:} \\\censor{\url{https://qped-eu.github.io/mass/\#/documentation}}}

Because of how Quarterfall handles submissions, two types of possible student submissions that are handled by MASS:
\begin{enumerate}
    \item Plain text via a text input field.
    This is usually done if the solution consists only of one class or of a number of methods.
    \item A ZIP-file containing the project as a whole.
    This is usually done if the assignment requires multiple classes.
\end{enumerate}

The configuration properties for individual feedback messages comprise:
\begin{itemize}
    \item \textbf{Message ID} Every message defined can have an optional ID.
    The ID can later be used to suppress different messages.
    An ID should be unique in order to avoid errors in the configuration.

    \item \textbf{Kind of Coverage Miss} Gives a choice between \textit{FULLY\_MISSED} and \textit{PARTIALLY\_MISSED}.
    The former designates that, to make the feedback message applicable, a line or a range of lines need to be fully missed, i.e., not even a part of it may have been executed.
    \textit{PARTIALLY\_MISSED} designates that a part of the line or a range has been missed but the line has been partly executed.
    For example, in Java, this can happen in lines containing an operator with shortcut evaluation like in \javaCode{a() && b()}: If the call to \javaCode{a()} returns \javaCode{false}, the result of the predicate would already be determined and the second method call would not be executed.
    If this is the only execution of that line, we have a partial miss.
    If the call to \javaCode{a()} returns \javaCode{true}, the second call has to be performed and there would be no miss at all.

    \item\textbf{File Name} The relative path to the file (not class) which needs to be tested within the project.
    It is important that the path from the package root is used.
    For example, following the Java convention to place classes in a \texttt{.java} file which resides in a sub-directory according to the package name of the class.

    \item \textbf{Line Ranges} Line ranges can consist of multiple lines or only one line.
    A line range is defined by either a line number or two line numbers (start to end).
    Multiple line ranges can be defined per message.

    \item \textbf{Message} The message that is shown when a miss is detected.
    It can be further customized by using markdown within the message.

    \item \textbf{Suppressed Messages} This field uses the aforementioned ID to implement a hierarchy of messages:
    If the currently configured message is applicable, it will suppress the message that have their ID listed in this field.
    Multiple messages can be listed.
\end{itemize}
\subsection{Minimal Example}
For illustration, let us consider the configuration for the minimal example from above.
We already established that two tests should be written: one for even numbers and one for odd numbers.
Furthermore, we can look at possible mistakes that a student can make and therefore configure a message for each possible mistake.
This means that we would need feedback in the following cases:
\begin{enumerate}
    \item the method is not tested at all
    \item only even numbers get tested
    \item only odd numbers get tested
    \item the test fails
\end{enumerate}
In our case, the feedback for 4) is automatically provided by the test suite and will be passed along to the student by default.
Because of this, we only need to concern ourselves with cases 1) - 3).
The messages for cases 2) and 3), however, should only be presented if case 1) is not applicable at the same time, because we do not want to give the solution away before the student has even tried to solve the task.
For this example, we will use the following messages in order of their respective feedback cases:
\begin{itemize}
    \item "You have not tested this method at all."
    \item "You should test for odd numbers as well."
    \item "You should test for even numbers as well."
\end{itemize}
Furthermore, to get the most out of the coverage checker, we have to provide an implementation of the class.
We can accomplish this by adding a link
\footnote{\censor{Private implementation of minimal example:}\\\censor{\url{https://qped-eu.github.io/mass/files/TeachersImplementation.zip}}}
to a zip archive within the configuration of MASS.
This is configured in the main form of the web app and is not shown here.
The file can, e.g., be provided via cloud storage with a private URL or using the file storage of Quarterfall.
A private implementation is necessary when we want to provide tailor-made feedback messages, as this is the only way that we can know the line numbers of the implementation of specific cases.

From figure \ref{fig:evenImplementation}, we can determine that the code that a student should write tests for is between lines 2 and 8 in the private implementation.
Furthermore, we can map our feedback cases to certain lines.
Not writing tests at all means that not a single line between 2 and 8 is covered, not even partially (case 1).
Only testing even numbers means that line 6 is not covered (case 2) and only testing odd numbers means that line 4 is not covered (case 3).

\begin{figure}
    \centering
    \includegraphics[width=\linewidth]{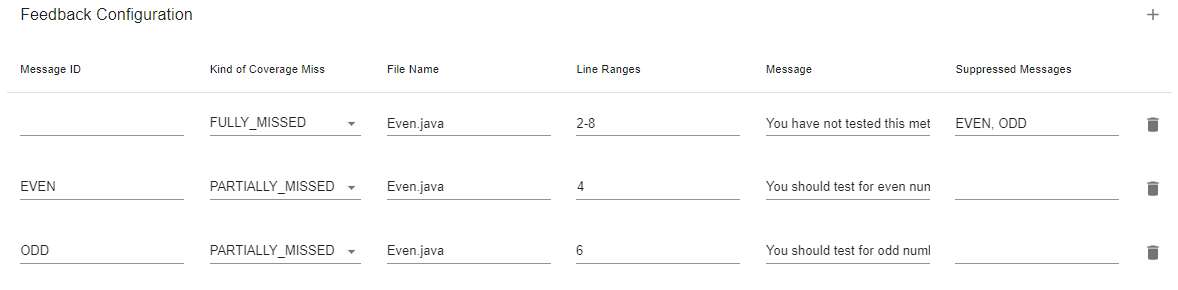}
    \caption{Configuration of the feedback messages for \texttt{Even}}
    \label{fig:evenFeedbackConfig}
\end{figure}

Let us discuss the different properties of the configuration for these three cases.
The complete configuration can be seen in Figure \ref{fig:evenFeedbackConfig}.
We have already decided that there should be a hierarchy of feedback messages such that case 1) suppresses cases 2) and 3).
Thus, we must give an \emph{ID} to the second and third message (first column in the figure).
In our example, we name the messages \textit{EVEN} and \textit{ODD} and enter these IDs into the \emph{Suppressed Messages} field of the first message.

Since case 1) encompasses the whole method, we need to select \textit{FULLY\_MISSED} for the \emph{Kind of Coverage Miss}.
Otherwise we would also trigger this message if the student tests the case for even numbers but forgets about odd numbers, in which case it would certainly be wrong to say that the method is ``not tested at all''.
The other two cases can be declared as \textit{PARTIALLY\_MISSED}.

The entirety of our private implementation, the execution of which we want to monitor, is in the file \textit{Even.java}, which is inside the default package.
Therefore, as \emph{File Name}, we specify this simple file name in all three cases.
In addition to the name of the file, we also need to specify the line ranges that the individual message needs to cover.
For the first message we need to specify lines 2 to 8 because that is where the whole method is located.
By the way, lines not containing executable code, such as lines with only an opening or closing brace, are ignored by our checker.
Thus it would have the same effect to specify, e.g. \texttt{3-6} or \texttt{3, 4, 6} for the \emph{Line Range}.
Cases 2 and 3 are covered in lines 6 and 4 respectively.
Next, for the \emph{Messages}, we simply insert the feedback messages presented above for the three cases.

\section{Integrated Example}

To demonstrate the procedural guidance as well as the MASS feedback toolkit in a realistic setting, we have developed an example assignment that integrates both.
It shows how the different steps of the procedural guidance should be mirrored in the chronological sub-questions of an assignment as well as the feedback that is presented while working on the assignment.
However, we will not be using the full Procedural Guidance in this integrated example.
Instead, the Procedural Guidance will be used by the teacher to let the students have the External Design as described in the Procedural Guidance.
In accordance with \cite{merrienboer2007}, a number of steps were produced by the teacher but the students are still required to fill in some blanks left open by the teacher.
The assignment is fully worked out in Quarterfall and we provide access to it as an interactive demo, as well as the full configuration data as an electronic appendix to this paper: \censor{\url{https://t.ly/b0V6}}.

The assignment starts with a description of the general goal, namely that students are supposed to develop an implementation of a \emph{bag}:
\begin{quote}
\itshape
In this exercise, you will implement a class \javaCode{collections.Bag} given a specification of this class. Furthermore, we ask you to implement a JUnit test class \javaCode{test.TestBag} according to the specification.
In mathematics, a bag (or multiset) is a generalization of the concept set.
An element of a bag can appear more than once in a bag, unlike a set where each element can appear only once. In this exercise, we limit ourselves to a bag of integers.
\end{quote}

Next, the students are provided with a specification (\emph{External View}) for the class \javaCode{Bag} as can be seen in Figure \ref{fig:bagSpec}.
The listing shows an excerpt of this specification, the full assignment (accessible online) is worked out for all seven methods as well as the constructor of the class.
The listing demonstrates how different cases are elaborated and conditions for different flows are logically grouped as contracts.
For example, there are two flows through the remove-method: the ``Happy Path'', when something actually gets removed, and the ``Non-Happy Path'', where the bag does not contain an item ready for removal.
The pre- and post-conditions for both flows are grouped logically, implying a certain sequential order for implementing the method.

\begin{figure}
    \centering
    \begin{java}
package collections;
/**
 * This class represents a bag of integers. ...
 * External invariant:
 * @invariant The number of elements in the bag
 *            is always >= 0
 */
public class Bag {
  /**
   * @desc Remove element elem from the bag
   * @param elem the element to remove
   * @return true if elem is removed from the bag
   *         otherwise, false */
  /*@ @contract Happy-path {
   @   @requires length > 0 and the bag contains n
   @             elements of elem, with n > 0
   @   @ensures new cardinality(elem) = old
   @            cardinality(elem) - 1
   @   @ensures length is old length minus 1
   @   @ensures \result = true
   @   @assignable elements
   @  }
   @  @contract Non-happy-path {
   @    @requires length = 0 or the bag does not
   @              contain element elem
   @    @ensures the bag is not changed
   @    @ensures length is old length
   @    @ensures \result = false
   @ }
   */
  public boolean remove(int elem) {
	  return false;
  }
  // ...
\end{java}
\caption{Specification of \texttt{Bag}}
    \label{fig:bagSpec}
\end{figure}

The first sub-question is as follows: 

\begin{quote}
\itshape
Based on the provided specification of class \javaCode{collections.Bag}, write a JUnit test file with all needed test cases.
Create a package \javaCode{test} with the JUnit Java class  \javaCode{TestBag}. 
\end{quote}

Generating feedback for this sub-question requires the strength of our coverage checker.
Particularly, students are asked to write their tests before starting the implementation of the \javaCode{Bag} class.
Normally, this would require them to use the stub implementation provided along with the specification in the assignment introduction and develop their tests against this, in order to arrive at a test implementation free of compiler errors.
They would, nevertheless, only be able to check the correctness of their test when the implementation of \javaCode{Bag} is ready.
And if both the test and the implementation contain mistakes that cancel each other out, it would go unnoticed.

With our approach, students can submit their test implementation before considering to fully implement \javaCode{Bag}.
For this purpose, we can configure the MASS checker.
In fact, the demo assignment reachable through the electronic appendix mentioned above also uses the syntax and style checker from the MASS toolkit.
Thus you can also get understandable feedback if submitted code cannot be compiled or coding conventions are violated.
For this paper, though, we will only discuss the configuration of the coverage checker.

Firstly, we need to provide a private implementation.
It should be noted that sometimes not all the different cases from the specification are reflected on separate lines in our code.
For example, some behavior depending on these different scenarios may be located in called libraries, and thus a single line would envelop multiple requirements.
For demonstration purposes, we have thus implemented the class \javaCode{Bag} by means of an \javaCode{ArrayList<Integer>} from the Java class library, which is stored in the field
\javaCode{elements}.
The listing in Figure \ref{fig:bagRemove} shows our implementation for the \javaCode{remove} method for which we have already seen the specification.
Other methods from the specification are implemented analogously.

\begin{figure}
    \centering
    \begin{java}
public boolean remove(int elem) {
  marker++; // You have not tested the remove method.
  if (elements.size() > 0 && elements.contains(elem)) {
    marker++; // You have not tested the requirement
              // `length > 0' and a bag containing elem
              // (happy-path scenario).
  }
  if (elements.size() == 0) {
    marker++; // You have not tested the requirement
              // `length = 0' (non-happy-path).
  }
  if (!elements.contains(elem)) {
    marker++; //You have not tested the requirement
              // `the bag does not contain element elem'
              // (non-happy path).
  }
  return elements.remove(Integer.valueOf(elem));
}
\end{java}
\caption{Implementation of \texttt{remove} in \texttt{Bag}}
    \label{fig:bagRemove}
\end{figure}

In order for the method to function correctly, only the implementation in Line 100 is needed.
But since we also want to check whether the method \texttt{remove} is invoked under the different circumstances given in the specification, we also insert code that is conditionally executed under these circumstances and whose execution can be observed by the coverage checker.

In the case of the \texttt{remove} method, there are two possible flows, namely the happy path and the non-happy path.
The precondition (``requires'') of the happy path contains a conjunction (``and''-operator) of conditions.
We map this to a single predicate in Line 86.
The coverage checker can only observe executable code, and while the \javaCode{if} statement on Line 86 will be executed, it will always be executed regardless of whether the condition holds or not.
We therefore also need to place an executable statement inside the \javaCode{if} statement that is only executed when the condition is satisfied.
What this statement does is irrelevant; we chose to increment the field \javaCode{marker}, which we have inserted for this purpose, as this is one of the simplest statements we could think of.

For the non-happy path, the precondition contains a disjunction (``or''-operator).
Ideally both possibilities for reaching the non-happy path should be executed in a test.
This is why the precondition is split into two conditional statements in our private implementation (lines 91 and 95), such that we can observe their execution separately.

The comments show the feedback message that we want to show when the corresponding line or code block is not covered.
It can be seen that the message refers back to the specification, partly by repeating the wording of the relevant case of the specification and partly by referring to the corresponding flow (``happy path'' or ``non-happy path'').

The marker code in Line 85 is executed unconditionally and, thus, indicates whether the remove-method has been executed during the test at all.
If this line is not covered, we do not want to further list any feedback messages that refer to specific parts of the specification of the method remove.
Instead we only want to point students to the fact that they failed to consider the specification of method \javaCode{remove} completely in their tests.
Thus, we configure that the feedback message corresponding to the miss of Line 85 suppresses the other feedback messages for method \javaCode{remove}.

\begin{figure}
    \centering
    \includegraphics[width=\linewidth]{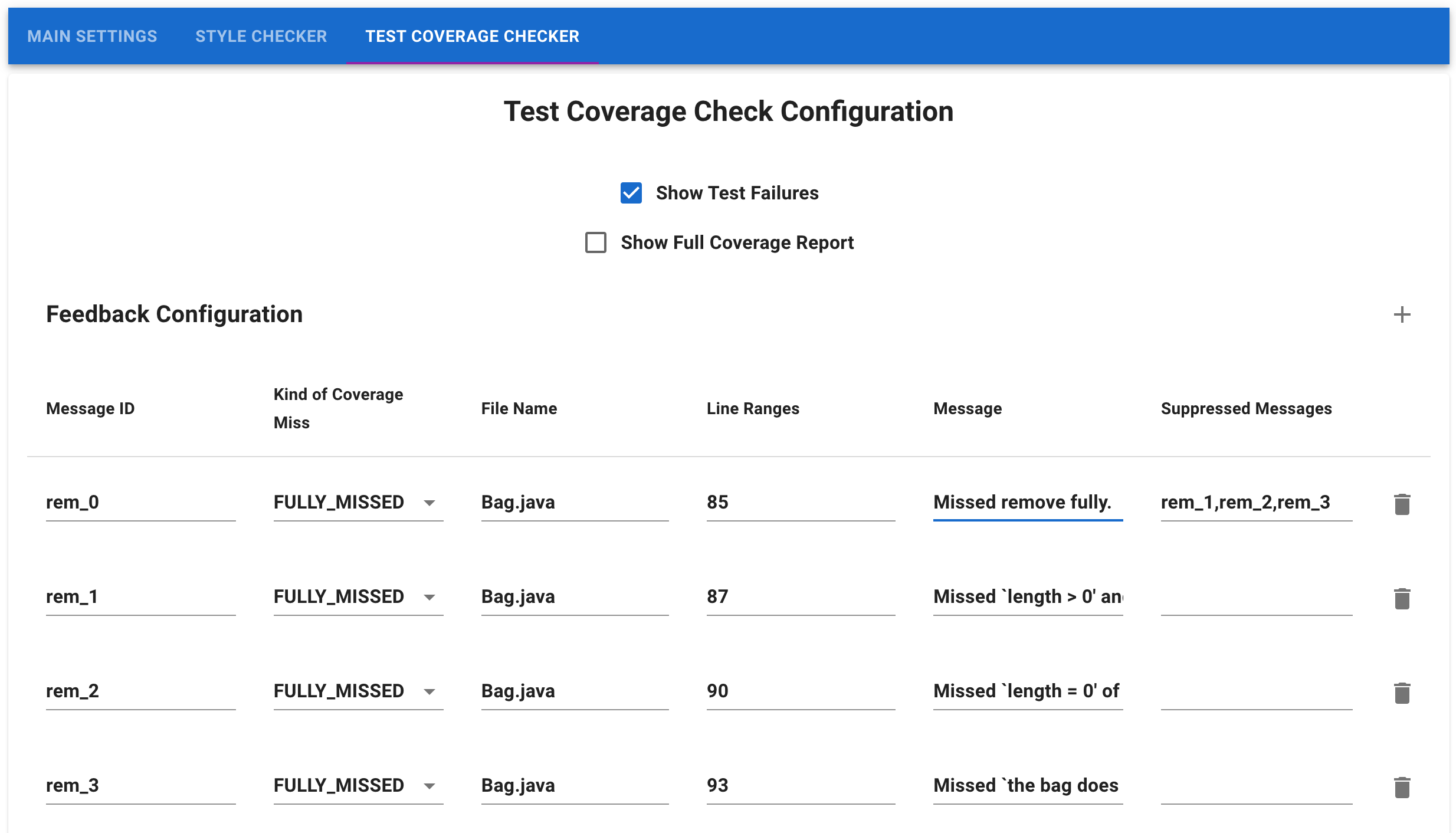}
    \caption{Configuration of the feedback messages for \texttt{Bag}}
    \label{fig:bagFeedbackConfig}
\end{figure}

An excerpt of the configuration for the test coverage checker can be seen in Figure \ref{fig:bagFeedbackConfig}.
The checkbox \emph{Show Test Failures} is also selected, which means that students do not only get feedback when cases from the specification are missing in their tests but also if their tests fail against the (presumably correct) private implementation.

In the second sub-question, students are supposed to do the next task of the procedural guidance, for which they receive a stub implementation containing the headers of all required methods together with the specification, which has already been seen in the assignment's introduction:

\begin{quote}
\itshape
Based on the specification of class \javaCode{collections.Bag}, implement all the methods.
\end{quote}

For this sub-question, the feedback generation is much simpler but can also be achieved by means of the test coverage checker.
In this case, students are supposed to provide the implementation of the \javaCode{Bag} class.
Although students implemented the test class in a test class in the previous sub-question, we cannot be sure if they really succeeded or maybe skipped the first sub-question; furthermore, it may still be the case that their test is incomplete or wrong in ways we did not foresee and therefore did not cover in the feedback configuration for the first sub-question.
We therefore provide our own, approved test implementation as a private implementation for the second sub-question, which can be seen in Figure \ref{fig:bagPrivateTest}.
In this way, we firstly make sure that the student implementation passes a complete and correct test suite.
Secondly, we can control the message to be displayed for failing assertions.
Again, we can refer to the relevant parts of the specification.
The configuration of the test coverage checker is trivial and only specifies that test failures are to be shown to students.

\begin{figure}
    \centering
    \begin{java}
@Test
void testRemoveHappyPath() {
   bag.add(1); bag.add(2); bag.add(2);
   boolean res = bag.remove(1);
   assertEquals(0, bag.cardinality(1),
       "The cardinality of elem 1 must be 0 "
       + "after the call remove(1) on the bag "
       + "{1, 2, 2}. (happy path)");
   assertEquals(3, bag.length(),
       "The length of the bag {1, 2, 2} after "
       + "the call remove(1) must be 3. "
       + "(happy path)");
   assertTrue(res,
       "The return value of the call bag.remove(1) "
       + "on the bag {1, 2, 2} must be true. "
       + "(happy path)");
}
\end{java}
\caption{Configuration of the feedback messages for \texttt{Bag}}
    \label{fig:bagPrivateTest}
\end{figure}

Lastly, students are requested to submit their full solution containing the implementation of the test \emph{and} the  \javaCode{Bag} class:

\begin{quote}
\itshape
Upload a zip file with your bag implementation and corresponding JUnit test file in it (both \texttt{.java} files). The ZIP file must have the following content:
 
\begin{verbatim}
+- collections/
|  \--Bag.java
\- test/
   \--TestBag.java
\end{verbatim}
\end{quote}

The feedback configuration for this assignment is just as trivial as that for the second sub-question.
The difference is that we do not provide a private implementation at all here.
Furthermore, in addition to checking the check box for showing all test failures to students, we also select the check box \emph{Show Full Coverage Report} in the web form.
This can be seen as a last sanity check before submitting their final solution:
As mentioned in the beginning, the MASS toolkit also checks whether the submission (the test and the bag class) contains any compiler errors or violates the coding conventions.
Furthermore, it checks whether the students' implementation violates one or more of their tests.
If any of these checks fail, students can go back to improving their solution.
They can also inspect the coverage report to see if there are untested parts of their implementation.
That may, e.g., be the case if they chose an unforeseen implementation approach which introduced additional distinctions of cases.
In that case, they can also go back to improving their solution, before making the final submission.

\section{Experiences}
In this section, we will discuss what we have achieved so far.
As we have already mentioned in the sections above, we have created a procedural guidance for students to follow in order to improve their understanding of test cases and a toolkit to provide automated feedback for their solutions considering the steps of the procedural guidance.
Thus, our approach for teaching students how to implement and think about testing is twofold:
We can use our procedural guidance to teach students on how to assess test cases and write tests at an earlier stage in developing their own software.
The resulting analysis can then be used by the teacher to configure and setup our toolkit to give students feedback on how well they followed the procedure and where they can still improve their understanding of writing test cases.

Both, the Procedural Guidance as well as our toolkit, are freely available under the MIT licence.
The MASS toolkit is available in our Git repository \censor{\url{https://github.com/qped-eu/MASS-checker}} and documented on our website \censor{\url{https://qped-eu.github.io/mass}}.
Our software toolkit passes a test suite of more than 200 test cases which are frequently executed by our continuous integration pipeline.
Furthermore, we have used our toolkit in a real life scenario in a beginner's course at our university.
During this run, we were able to identify two main points of improvement that we will be encompassing in future releases.

In the early phase of using the toolkit, we encountered that the time between requesting feedback and actually receiving feedback was perceived as too long; the performance is limited since the execution on the Quarterfall server happens in a resource constrained environment.
It should also be mentioned that the efficiency is limited by the use of JUnit and the JaCoCo coverage analysis tool.
We nevertheless took the criticism about the performance to heart and optimized our toolkit for efficiency.
Now students only have to wait approximately ten to 15 seconds when requesting feedback, which we consider an acceptable delay in the process of working on an assignment.
This is especially true when compared to the response time of a human mentor, which can rather be measured in minutes in the very best of cases.

Another common response was that it is quite cumbersome for students to develop their solutions within an IDE only to then upload the solutions to a platform, rework their solutions and repeat the process.
To address this criticism, we are developing a plug-in for both, IntelliJ IDEA and Eclipse, that directly communicates with and uploads the submission to a server running the open source Quarterfall server.
The plug-in will then receive the feedback generated by MASS and pass it on to the students without the need to visit an external website.

To ensure a good usability for teachers, we have developed and published a website with documentation, the configuration tool and tutorials to guide teachers through the process of using and configuring MASS.
There is a tutorial present for every checker implemented in MASS that guides a teacher through the process of configuring that checker to their liking and their needs.
The same website also hosts the graphical user interface for configuring every checker that is currently implemented within MASS.
As we have already discussed, MASS uses a specific JSON-object as an interface to communicate with any front-end that adheres to the same configuration as our interface.
The graphical user interface can be used to easily generate a valid JSON-object for the configuration of our checkers.
The code for our website and by extension also our configurator is published open-source on our Github-repositories.

Working with MASS during the aforementioned course, we gathered data and published our findings in \censorCite{dick2024} to evaluate the usage of the MASS toolkit.
To recap our findings from this prior publication, we added a diagnostic task to the final exam of the course where we introduced MASS in.
This diagnostic task was added in three consecutive runs of the course.
In the first run, we did not introduce any new concepts into the course to get a baseline to measure against.
In the second run, we introduced the TILE approach as proposed by Dorn, et al \cite{tilePaper2023}.
Lastly, in the third run, we introduced MASS to our students.
Using our diagnostic task, we were able to see that our students' ability in our diagnostic task increased with statistical significance.
This significant increase has been shown from the first run to the third run.
Because of this, we can state that MASS and TILE have both contributed to the statistically significant increase in our students' performance.

We have some prior experience with procedural guidance, namely in teaching multi-threaded programming to more advanced students, reported on in \censorCite{Bij2017} and \censorCite{Bij2019}. 
An important insight was that providing step-wise guidance was helpful and appreciated by students.
However, applying the explicit but abstract steps, formulated in general terms, to actual assignments was still not easy for students.

In the present research, we attempt to remedy this by providing more structure that has more detailed steps for the students' activities at a high level. 
More importantly, we now provide guidance-for-the-guidance at the assignment level:
In the assignment, specific demands about the artefacts that should be produced in the solution are made.
Furthermore, we provide tooling that can assess solutions and give feedback on the quality of that solutions:
The test tool is our most advanced contribution to this.

A more detailed evaluation paper is in preparation. 
An initial insight, obtained during the application of the approach, is that not requiring students to elicit a specification, but rather providing it, was experienced by the student as very useful. Following Kirschner~\cite{kirschner2008ten}, we gave the students the design and specifications and ask them 1) to implement the functional code and 2) to design and implement the corresponding tests, both tasks supported by the procedural guidance. The quality of the artifacts in terms of correctness and readability improved significantly: the percentage of errors in the code dropped significantly (about 80 percent) and the tests found many more errors. Subsequently, students can gradually be tasked with developing portions of the specifications, fostering iterative learning and skill development.

\section{Limitations and Future Work}
One improvement we identified in our current implementation is that the feedback is static between attempts.
As it stands, it is not possible to grant a student further insight depending on the number of attempts that they used to arrive at their current solution.
This would be beneficial to students struggling with a certain task.
The Quarterfall platform already supports such a behavior, as along with the student solution it also passes the number of the current attempt to the feedback tool.

However, there are also challenges needing to be solved.
First of all, we need to identify in which way the feedback should change with an increasing amount of attempts.
For example, the messages can change (e.g., becoming more directive over time) or the ``suppression'' of feedback messages can stop.
Second, a system that solely counts attempts could be easily abused to get the most directive feedback nearly instantly by submitting empty files or the same files over and over again.
Or, more accidentally, if students are supposed to write tests for multiple methods, students could get stuck on writing a test for one method, needing multiple attempts.
Then they should ideally not be presented with more directive feedback on testing the methods which they have not yet considered.
Thus, a system for dynamically adaptive feedback would need further research to identify relevant cases and easy-to-use ways for configuring them as well as consideration and fine-tuning in order to prevent students from simply abusing the system and to prevent the system from accidentally giving away too much information.

Another point of improvement is the configuration of feedback.
Naturally, configuring our tools using only the native JSON-object is a huge and cumbersome undertaking.
This is somewhat alleviated by using our configurator on the website.
We've also developed a tool that uses comments within source code files to build the configuration for the test checker.
The teacher would only need to annotate the solution, upload the project as a zip-file to our website and the configuration will be done based on the annotations.


Lastly, maximizing the code coverage does not necessarily lead to good test quality.
For example, Edwards et al. \cite{EDWARDS2014} found that students can reach a high percentage of coverage without finding many bugs.
They found that students could cover 95\% of their own solution while only finding about 14\% of faults within their solution.
We also see this limitation in our approach of using test coverage analysis, as we can only observe whether code in the known (private) teacher implementation is executed or not.
By using the approach presented in the previous section, specifically including code in there that is only conditionally executed depending on the argument values, we can check whether the tests of the students cover the different cases of the precondition of a method that is to be tested.
What we cannot check, however, is whether the test cases contain assertions that cover the different cases of post-conditions.
For example, assume that a test for the Bag-assignment only asserts that the \javaCode{size()} method returns a value $\geq 0$; a correct private implementation would certainly pass this test, but it is also obvious that the test is not sufficient.
Mutation testing \cite{mutation} is an existing approach for assessing the strength of test suites and would be able to discover such a weakness.
It works by injecting simple mutations into the code which presumably introduce errors in the code, and determining whether tests start failing after the mutation.
When this is not the case, the test cases are too weak, as they do not detect the injected errors.
We are therefore researching mutation testing or variants thereof as a means for extending our test-coverage checker.
Our system could then take the result of the mutation testing and give feedback either customized by the teacher or generic feedback in the cases where a teacher would not have set-up custom messages to the student.


\section{Related Work}\todo{The bibliography is very compact and some text passages with a subsequent citation are difficult to link to the reference they point to.}
When talking about automated feedback generation in the programming space, it is crucial to look at different systems that are on the market.
Some universities have developed their own solution over the years to help their students improve upon their solutions without further human interaction.
Mostly, these systems are used to grade students and assess their solutions based on static and dynamic analysis.
This is mostly done by using a compiler and predefined unit-tests.
We will include some of the more well known tools in this section.

One such system for giving automatically generated feedback is INLOOP, the INteractive Learning center for Object-Oriented Programming, \cite{morgenstern2018continuous} developed by the university of Dresden.
INLOOP uses a push and pull based environment which supports any programming language that can be used in a docker-container.
It also has the ability to use predefined Unit-tests to test a student's solution.
The feedback provided by INLOOP is constricted to the name of the predefined tests and their current status, passed or failed.

Another system for automatically generated feedback is the system JACK by the university of Duisburg \cite{goedicke2008computer}, which supports several different kinds of tasks.
In JACK programming tasks are solved locally and then solutions are uploaded to the server, which gives feedback based on predefined unit tests.

EMSEL is a tool that is used for teaching HTML and JavaScript \cite{EMSEL}.
Though it is made for HTML and JavaScript, it can easily be expanded to other programming languages.
As a teacher, it is possible to define programming tasks and their solution.
A student would then need to upload their solution and receive feedback for improving themselves.

ArTEMiS is also a tool for receiving instantly generated feedback on programming tasks \cite{artemis}.
It is independent of programming languages and uses an online editor for handing in assignments as well as the possibility to use a git-repository that is setup by ArTEMiS.
After submitting a solution, it is then analyzed by ArTEMiS by using predefined Unit-tests.
These predefined tests are then used to give a student feedback in the form of passed and failed tests as well as the generic message of the test suite.
Importantly, the used test cases are not visible to students to prevent them adapting their code to suit the tests.

All of these systems considerably differ from MASS, which itself is designed to be used in conjunction with Quarterfall but is inherently system-agnostic.
It has the ability to customize feedback messages not only for tests but also for style violations, syntax errors and more.
Generic feedback messages can also be tailored to the level of knowledge of students so they will not be overwhelmed with unknown words or concepts.

A recent and more complete overview of feedback tools is provided by Keuning et al. \cite{keuning2018} who analyzed and categorized the feedback generation of over 100 feedback tools in a comprehensive study.
They found that the tools mostly focus on identifying errors in student submissions and telling students where their mistakes lie without giving them feedback on how to fix their mistakes in the future.
Furthermore, they found that in general the feedback can not be customized by the teacher beyond providing input data or unit tests.
Because of this, students at different knowledge levels receive the same feedback which could lead to confusion in beginners when they see advanced technical terms that they would not yet be able to understand or advanced students that get bothered by feedback aimed at beginner-level students without advanced technical terms that they would expect in such feedback.
With MASS, we aim to address all of the points mentioned by Keuning et al., specifically the quality and customisation of feedback to suit the needs of the students at that specific point in their education.
\todo{related work hinzufügen}

\section{Conclusion}
Over the course of this paper, we have presented our approach for teaching students to develop software together with relevant and complete tests, based on procedural guidance.
By describing informal rules of how to arrive at each stage coming from the previous one, we provide a lightweight, yet very efficient guidance to students.
Making test development an explicit step and giving students rules for creating their tests improves their willingness to develop tests.

We have also shown our feedback toolkit called MASS, which can provide students with automatic and individual feedback on programming assignments.
It is designed to be integrated into the web-based e-learning platform Quarterfall, but can also be used independently.
MASS provides checkers to produce feedback for different aspects of programming assignments, including syntax, style and test coverage.
Furthermore, it is designed to let teachers configure the feedback at a fine-granularity per assignment in such a way that the knowledge level of the students at the current time as well as peculiarities of the assignment can be reflected.
The test coverage checker is the most advanced one and presented in much detail.
MASS also allows teachers to specify custom rules for the completeness of student tests based on code coverage and custom messages for missing test cases.

In an integrated example, we have fully worked how the procedural guidance and MASS can intersect to help students with their solutions.
Within this example, we have shown how the procedural guidance is used to help students arrive at their solution and how the configuration for the test coverage checker can be derived from the specifications used in the procedural guidance.
Furthermore, we have shown in an earlier publication that using MASS does indeed benefit students in writing better test cases and have shared preliminary results of a study on the usefulness of the procedural guidance presented within this paper.

Because MASS provides an easy way for automatically generating feedback on the completeness of test cases, we expect that it will become more common to consider the quality of tests in student solutions throughout programming courses.
Commenting regularly on the availability and completeness of tests in their solutions, even if the main objective of the assignment is different, will demonstrate the importance of testing to students and stimulate them to put more effort into writing tests, which will finally lead to them writing better programs.
        
\end{document}